\documentclass[conference]{IEEEtran}
\IEEEoverridecommandlockouts
\usepackage{cite}
\usepackage{amsmath,amssymb,amsfonts}
\usepackage{algorithmic}
\usepackage{graphicx}
\usepackage{textcomp}
\usepackage{xcolor}
\usepackage{enumitem}
\usepackage{wrapfig}
\def\BibTeX{{\rm B\kern-.05em{\sc i\kern-.025em b}\kern-.08em
    T\kern-.1667em\lower.7ex\hbox{E}\kern-.125emX}}
    
\newcommand{\awesome}{\textsc{Awesome}~}
\newcommand{\boutique}{\textsc{Boutique}~}

\begin{document}

\title{Multi-Model Investigative Exploration of Social Media Data with \boutique: A Case Study in Public Health\\
}

\author{
\IEEEauthorblockN{Junan Guo}
\IEEEauthorblockA{\textit{Dept. of Mathematics} \\
\textit{Univ. of California San Diego}\\
La Jolla, USA \\
jug016@ucsd.edu}
\and
\IEEEauthorblockN{Subhasis Dasgupta}
\IEEEauthorblockA{\textit{San Diego Supercomputer Center} \\
\textit{Univ. of California San Diego}\\
La Jolla, USA \\
sudasgupta@ucsd.edu}
\and
\IEEEauthorblockN{Amarnath Gupta}
\IEEEauthorblockA{\textit{San Diego Supercomputer Center} \\
\textit{Univ. of California San Diego}\\
La Jolla, USA \\
a1gupta@ucsd.edu}

}

\maketitle

\begin{abstract}
We present our experience with a data science problem in Public Health, where researchers use social media (Twitter) to determine whether the public shows awareness of HIV prevention measures offered by Public Health campaigns. To help the researcher, we develop a \textit{investigative exploration} system called \boutique that allows a user to perform a multi-step visualization and exploration of data through a dashboard interface. Unique features of \boutique includes its ability to handle heterogeneous types of data provided by a polystore, and its ability to use computation as part of the investigative exploration process. In this paper, we present the design of the \boutique middleware and walk through an investigation process for a real-life problem. 
\end{abstract}


\section{Introduction}

For a society to be healthy, it is not only important that individuals live a healthy lifestyle, it is equally necessary for them to be aware of diseases they may be susceptible to, their risk factors, and preventive and treatment options available to them. The task of public health researchers and practitioners is to be vigilant and ensure the conditions in which people can be healthy, prevent people from getting sick or injured, and conduct research to determine the causes and remedies of events like epidemiological outbreaks. 

Traditionally, the instruments of monitoring and measuring the state of Public Health are public surveys and compiled data from hospitals and clinics. This data is often of high quality, but suffers from a lack of scalability because surveys are expensive, often focused on a narrow sub-population, and are conducted in with a limited time-scope. More recently, researchers in Public Health are increasingly turning to the social media as an alternative information source for tasks like estimation of the size and location of an epidemic, or the public awareness of a government-sponsored treatment method \cite{young2013social, bychkov2018social}. The expectation is that the social media will help access a broader cross-section of the desired population over a more extended period of time than surveys. In this paper, we present a case study developed for the benefit of Public Health researchers who study HIV/AIDS, and investigate progressively-defined research questions like the following: \begin{itemize}[leftmargin=*]
    \item Are people, who are at a high risk for HIV, aware of the preventive measures available to them? (\textbf{R1})  
    \item  How are they becoming aware of these measures? (\textbf{R2})
\end{itemize}
Our goal here is not to answer these questions, but to provide a mechanism for public health researchers to formulate questions like these through  exploratory analysis of continuously monitored social media. 

\noindent\textbf{Investigative Exploration.} We posit that a tool that addresses the domain science requirements above needs more than the standard information exploration mechanisms offered by today's data platforms \cite{bradel2015big,siddiqui2016effortless,di2017exploratory,anagnostou2017alpine}. 1) While a standard data exploration system typically handles a single type of data (e.g., relational), the desired system should allow exploration of \textit{heterogeneous data collections}. Social Media data is considered \textit{heterogeneous} because it can be segmented to subsets of data, where each subset may belong to a different data model. In the case of \boutique, which builds on the \awesome polystore platform \cite{gupta:bigdata:2016, DBLP:conf/bigdataconf/DasguptaMG17} where the social media is represented as a combination of relational, graph and text data where each type of data can be temporal. 2) It should enable the analyst to start with a simple operation like a keyword search (e.g., which users have at least $k$ tweets on HIV or AIDS) and data browsing (e.g., sample tweets of the above group), but then then progressively deepens the enquiry through a series of retrieval \textit{and} analytics tasks before deciding on the next information seeking step.  3) At each  exploration step, a user may choose to perform additional computation, and decide to store aside the results of the computation, which can be used in a subsequent step of the exploration; 4) The user should be able to ``backtrack'' to a prior step in the exploration, and yet be able to re-use the results of any computation that has already be performed on the data. We use the term \textit{investigative exploration} to refer to this form of iterative combination of exploration and analytics. 

In the next  section, we present the design of the \boutique system and in the next section we describe how the Public Health questions can be formulated and answered through it.

\section{The Design of the \boutique System}
\label{sec:arch}
\begin{figure}[ht]
    \centering
    \includegraphics[width=0.5\textwidth]{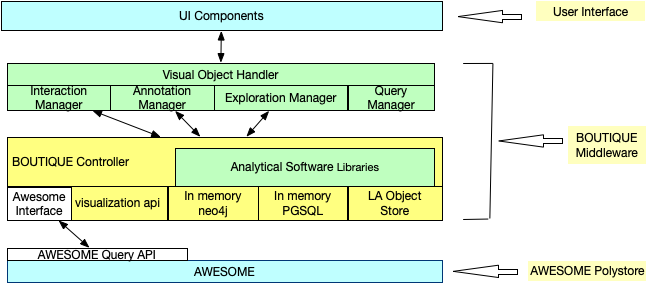}
    \caption{Functional architecture of the \boutique system}
    \label{fig:arch}
\end{figure}

\noindent Figure \ref{fig:arch} shows the architecture of the \boutique system, which is designed as a middleware platform that on the one hand queries and receives data from the \awesome polystore, and on the other hand, interacts with the user by presenting data through a set of visual objects. A user can (i) interrogate a visual object (e.g., a hashtag histogram), (ii) manipulate its content, (iii) invoke further analyses from it, and (iv) launch a new visual object based on the \textit{exploration state} of the current visual object. 
These computed objects are stored in in-memory data and can be reused throughout/ the exploration session. 
The components of Figure \ref{fig:arch} are explained below.

\noindent\textbf{Multi-Model Materialized Views.} The \boutique system recognizes the following types of data objects -- relations, property graphs (graphs whose nodes and edges can have properties represented as attribute-value pairs), document objects like JSON and XML, and ``linear algebra objects'' like a vector or a matrix. The system administrator of a \boutique application defines a set of materialized views that query the \awesome platform and constructs a corresponding data object in the middleware. The materialized views capture aggregate data (e.g., a hashtag co-occurrence statistics), as well as samples of actual data records that users can browse.


\noindent Materialized views are stored at the polystore (for refreshing) and in the \boutique layer, are ingested into (a) in-memory data stores for relational and graph data and (b) in-memory data structures for linear algebra data. The in-memory destination for a materialized view depends on the model the view subscribes to -- if a view materialized from a graph store has a tabular structure, it is stored as a relation in the middleware. To maintain consistency in the middleware all materialized views are updated synchronously -- however, at this point of time, the refresh is scheduled by the system designer and the views are not incrementally maintained at level of the middleware.

\noindent\textbf{Visual Objects.} A visual object is represented as a 6-tuple $V=(visID,vType,dType,\overrightarrow{parameters},graphic,state)$. $visID$, the visualization identifier, is unique within an exploration session, $dType$ refers to the data model of the content that will be visualized, $vType$ is one of the system recognized visualization types (pieChart, heatmap, multiTimePlot, labeledGraph $\ldots$), $\overrightarrow{parameters}$ is the actual data used for visualization, and $graphic$ is the visual item displayed. The actual specification of $\overrightarrow{parameters}$ varies depending on $dType$ and $vType$. For example, for a relational table, two categorical columns, together with a numeric aggregate column can be used to create a heatmap -- the parameters identify the columns, and point to a configuration file that has the visual parameters to create the $graphic$ object. Every recognized $vType$ has a set of \textit{interaction methods} that specify the operations that can be performed on the visual object. The $state$ item contains  current values of the interaction parameters. For a multiTimePlot (where the X axis is time and the Y axis plots the values of one or more categorical variables like a hashtag as they change with time), the $state$ parameters can be a user-selected time interval or a set of user-selected Y axis values (e.g., hashtags). A visual object is paired with a specific in-memory data object created through materialization to instantiate the visual object. 
All visual objects used in a session are maintained by \boutique in case the user returns to a previous visualization state. Since one can navigate from one visualization to another, the transition information is represented as a tree to help navigation.

\noindent\textbf{Interaction Management.} A user interaction with a visual object is some combination of a visual operation, a computation request and a query. A user can, for instance, choose a time interval on the timeplot of tweet count, and then initiate a computation/query to detect hashtags that show a ``bursting'' behavior in the selected time-period. This involves computing the value of a metric called \textit{local peaky-ness} on the time series (of total tweet count), gather the smaller time intervals in which any bursty behavior might have occurred, and pass the intervals as parameters to a top-$k$ query that retrieves the hashtags that show a bursty behavior within those time slots. The interaction manager 1) formulates unambiguous queries and 2) determines the right representation for a visual object.\\ 
\noindent 1) Suppose a user selects an interval on a timeline, and computes all hashtags that burst during the interval. Next, the user asks to see a union of the hashtag co-occurrence network and the mention-network of users who used those hashtags at least 10 times. First, the interaction manager allows the user to formulate cross-visualization queries (see Section \ref{sec:exploration}). In doing so, it discovers potential ambiguities. In our example, the ambiguity in formulating the network query arises because it is not obvious if the user-selected interval from the timeline object should be passed to the network retrieval request; hence, it asks the user to resolve the identified conflict.\\
\noindent 2) In many cases, a visual object can be displayed in many ways. For example, a histogram can be represented as table, a bar chart or a pie chart. The \boutique system implements and extends the representation scoring method in DeepEye \cite{qin2018deepeye}. This scoring method evaluates attribute pairs based on their data properties (categorical vs. ordinal vs. continuous, temporal vs. static, number of discrete values $\ldots$) and assigns scores to possible visualizations that users usually find visually satisfying. The highest ranked visual representations are then selected to display the result of a visual computation.


\noindent\textbf{Annotation Management.} An annotation is a piece of information generated due to a user action and is expressed as a relation associated with a specific visual object. A user can associate a set $\mathcal{A}(visID)$ of annotations with a single $visID$, which are considered to be AND-ed. The possible set of relations and their argument types is fixed for every $vType$. For example, for a multiTimePlot visual object, the assertion can be about a specific time interval, a specific categorical variable on the Y axis, a specific value at a certain time, etc. For example, if a user selects a sub-interval from the time plot and gives it a label, then the visual object, the sub-interval and the label form an annotation that a user may opt to store. 

\noindent One use of an annotation is to pass variables from one visual object to another. Suppose a visualization $V_1$ produces the annotation set $\mathcal{A}(V_1)$, and the user navigates from $V_1$ to visualization $V_2$, then annotations from  $\mathcal{A}(V_1)$ can be used in $V_2$ if $V_1$ and $V_2$ have common variables. Suppose $V_1$ is the timeline that plots retweet/tweet ratio over time for hashtags $H_1 \ldots H_{10}$, and $V_2$ is a reverse sorted bar chart of significant non-hashtag terms for a time interval selected by the user in $V_1$. Then, one can launch $V_2$ from $V_1$ by annotating and storing hashtags from $V_1$.

\noindent \textbf{Analytical Operations.} The \boutique system is integrated with a number of analytical libraries so that users can use the data selected from any visual object to a perform analytical operations. For example, a frequency distribution can be tested for its fit to a model, or two graph objects can first be individually analyzed to compute their node centrailities, and then the centrality distribution can be compared to test for their similarity. Tasks like model training are harder to perform within a visual exploration dashboard. However, a user may select training data from the dashboard and ship it to a Jupyter Notebook where a user can develop a training model (e.g., for user classification). Once trained, the model can be brought back to add to the user's private model library so that new unseen data can be sent to the model directly from the dashboard. The results of the analysis (e.g, the class assignment created based on the user's model) creates a new data attribute -- this is treated as an annotation and can be added to the \boutique system. If the user opts to permanently store the results of an analysis, \boutique updates the \awesome system with the new information.

\noindent\textbf{Query Processing.} Queries against materialized views are issued when a new visualization is created and during a data manipulation operation performed in the focus object. A query may invoke any analytical function registered with the \boutique system. \boutique uses a set of pre-defined query templates associated with each visualization type (including a Lucene-style search template). A template implements a fixed query plan over the in-memory databases and data structures, and assembles results into a data format required by the initiating visual object. The query processor uses a number of cross-model operations like merging graphs with relational data as well as performing similarity join between tweet and news content based on entities represented by news terms and hashtags respectively. Several cross model operations use join indexes. The first index maps edge objects in the graph store to records in the relational store. The second index maps dense subgraphs of a graph to term vectors that can be used to query an in-memory inverted index stored as a materialized data structure. The third cross index is implemented as a time hierarchy such that temporal data of different granularity are indexed through this tree.

\begin{figure*}[ht]
    \centering
    \includegraphics[width=0.75\textwidth, scale=0.9]{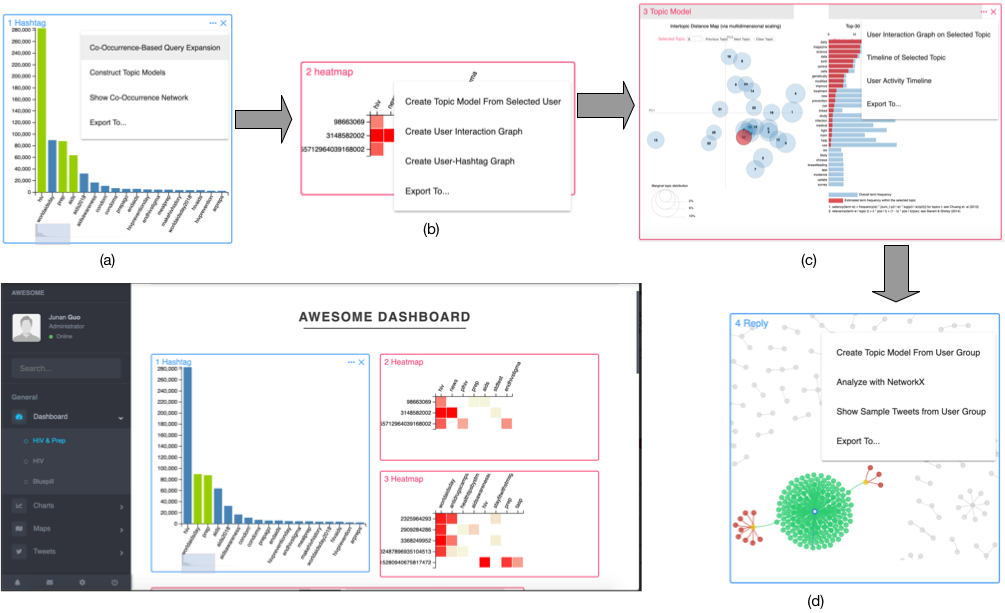}
    \caption{The \boutique dashboard. (a) Hashtags are selected (green) from a bar chart; (b) User seeks a heatmap from selected hashtags ; (c) A topic model of selected users from heatmap; (d) A network neighborhod selected for NetworkX analysis. (center) The user selects two heatmaps for comparative analysis }
    \label{fig:interface}
    \vspace{-1em}
\end{figure*}

\section{The Process of Investigative Exploration }
\label{sec:exploration}
We use Twitter as our data source. For this problem, the Twitter stream is filtered with a keyword list provided by the domain scientists. To use the \boutique infrastructure to answer questions \textbf{R1} and \textbf{R2}, a user will reformulate the questions in terms of the operations offered by the system using a human-in-the-loop analysis. Our exploration strategy is can be laid out as follows. (a) First identify the topics (represented as hashtags) of conversation of users who mention the terms HIV or AIDS more the $k$ times; (b) using these topics, identify tweet authors who, by human judgment, belong to the at-risk community; (c) analyze the topic distribution of these users to determine whether they converse about HIV/AIDS prevention measures; (d) determine the interaction network of the users who participate in such conversations and their HIV related topics; (e) finally, determine if there are dominant users who influence the spread of the HIV-prevention related messages in this network. Fig. \ref{fig:interface} shows the steps executed in the \boutique dashboard.

\noindent \textit{Step (a)}: The user performs a hashtag search on (HIV OR AIDS) and seeks a histogram of hashtags from tweets that use either of these hashtags. The system presents the top 100 hashtags in the form of a bar chart (due to the DeepEye algorithm), from which the user interactively suppresses non-informative hashtags (e.g., \#hiv), As hashtags are removed from the interface, the system updates the visual object so that the now-top-100 are visible. The user visually selects relevant hashtags (e.g., \#condomless) that are deemed important.

\noindent \textit{Step (b)}: Although the user has selected has some hashtags in Step (a), they are clearly not complete because there may be other tweets with other hashtags that did not show up in that step, and yet are relevant. This is done by a \textit{co-occurrence-based query expansion} process where the top $n$ hashtags that most frequently co-occur with the selected set are also used. To perform this operation, the user just seeks a tweet distribution by authors, turns on the query expansion flag. The system automatically generates a heatmap of hashtag vs. number of tweet-authors. The number of results is chosen by default, but the user can ask for more results. The user may choose to select some or all of the users from the heatmap.

\noindent \textit{Step (c)}: To get the topic distribution from all tweets by users identified in Step (b), user invokes a topic model computation which is performed using the PyLDA library integrated with \boutique. The hyperparameter for this computation is the number of topics, which is chosen by default, but can be set by the user. The topic visualizer produces three outputs -- the terms associated with each topic, the tweets that primarily contribute to each topic, and an inter-topic correlation. The user inspects these topics to determine a correlated subset of topics that relate to HIV prevention, and preserve the topics as annotations to be used in the next step.

\noindent \textit{Step (d)}: From the topics chosen in Step (c), the user would like to construct a network whose nodes are topic terms and users. The edges are standard social media relationships like \texttt{mentions, follows} and \texttt{replies-to}. In \boutique, every visual object involving two or more entities (e.g., tweet-authors and hashtags) offers to construct a network where the investigating user tells the system what network to create. The network is created as a property graph that can be queried using OpenCypher \cite{green2018opencypher} and analyzed using graph algorithms provided by the Python library called NetworkX. In contrast to Steps (b) and (c), which involves direct query of data stored in \boutique, Step (d) requires a \textit{model transformation} operation whereby relational data from twitter is converted into graph data, which is now stored in the system.

\noindent \textit{Step (e)}: Once a graph is constructed and visually presented, the user can perform visual operations like neighborhood selection, eliminating nodes/edges from display and requesting computations on selected subgraphs. For our analysis, the user selects subgraphs induced by tweet-author nodes and ask to compute an influencer detection operation, which currently finds nodes that have high PageRank and high betweenness centrality. These operations are performed by first converting the property graph to a NetworkX object and then invoking NetworkX to compute the two centrality scores. The resulting data is filtered by the query processor to return the desired set of tweet authors and their reachability graphs to the user.
\section{Conclusion and Future Steps}
\label{sec:conclusion}
In this paper, we presented some initial results from our experience in analyzing social media data for a real-life Public Health problem. One goal of this study was to determine if the \boutique dashboard and its operations provide the right level of abstraction for a data scientist in Public Health. The design of the system has been strongly influenced by their positive feedback. Our future work is to develop more functionality like cohort comparison and locational analysis into the system.\\
\textbf{Acknowledgment.} We are grateful to our collaborators Drs. Sean Young and Wei Wang (UCLA) and Dr. Nanette Benbrow (Northwestern University) for their suggestions and advice.

\bibliographystyle{abbrv}
\bibliography{boutique}
\end{document}